\documentstyle[preprint,aps]{revtex}
\begin{document}
\draft
\title
{Quenched and negative Hall effect in periodic media: \\
application to antidot superlattices}
\author{R. Fleischmann$^1$, T. Geisel$^1$, and
R. Ketzmerick$^{1,2}$}
\address{
($\;^1$) Institut f\"ur Theoretische Physik und
Sonderforschungsbereich Nichtlineare Dynamik, \\
Universit\"at Frankfurt, D-60054 Frankfurt/Main, Federal Republic of Germany\\
($\;^2$) Physics Department, University of California, Santa Barbara, CA 93106,
USA}
\maketitle

\begin{abstract}
We find the counterintuitive result that electrons move in {\em opposite}
direction to the free electron ${\bf E \times B}$-drift when subject to a
two-dimensional periodic potential. We show that this phenomenon arises from
chaotic channeling trajectories and by a subtle mechanism leads to a
{\em negative} value of the Hall resistivity for small magnetic fields.
The effect is present also in experimentally recorded Hall curves
in antidot arrays on semiconductor heterojunctions but so far has
remained unexplained.
\end{abstract}
\narrowtext

We have long become used to the fact, that the Hall resistance of
{\em extended} systems grows linearly with the magnetic field $B$ apart
from quantum phenomena like the quantum Hall effect. It thus came
as a surprise when for small $B$ in antidot superlattices the Hall
resistance remained quenched and even took on negative values
\cite{weiss,ensslin} although the current is carried by electrons and
not by holes. In {\em spatially constricted} systems like ballistic
microjunctions of crossed wires similar phenomena had previously been reported
\cite{roukes,ford,hansen} and were attributed to ballistic rebound
trajectories, where after colliding with a confining wall
the electron is ejected into the opposite lead
\cite{ford,been,baranger,cross}. Since this explanation rests upon
the finiteness and smallness of the junction, it cannot account for
large and extended systems. Thus the negative Hall resistance in antidot
superlattices (with as many as $10^{5}$ lattice cells) has remained
mysterious and one may have wondered whether it is not an artifact or an
edge effect.

In the present paper we describe a general mechanism whereby the presence
of a periodic superlattice potential causes a majority of electrons to move
in a direction opposite to the free electron ${\bf E \times B}$-drift.
It gives rise to quenching and negative Hall resistivity as a {\em bulk}
property. In the regime of small magnetic fields there is a long-time tail
associated with chaotic {\em channeling} trajectories, which tend to follow
one of the channels of the 2D potential as they are trapped close to a tube
of regular motion in phase space.
The escape from these channels after crossing 3 to 1000 or more
lattice cells is predominantly in opposite direction to the Lorentz force.
The latter phenomenon bears a partial resemblance to the
rebound trajectories responsible for quenching in ballistic microjunctions,
where in a sense one deals with a single unit cell of an antidot lattice.
However, it is the combination of the long-time tail and the asymmetric
escape that gives rise to the negative Hall effect in periodic media,
as one can see from two arguments.
Rebound trajectories can have a larger curvature
(radius $\approx a$) and would occur up to much higher magnetic fields
($B/B_{0}\approx 0.5$ in Fig.~\ref{rxy}), where quenching would be
predicted but has disappeared in our extended system. This is the regime
of magnetoresistance ($\varrho_{xx}$) oscillations due to {\em cyclotron-type}
orbits \cite{weiss,prl} where the Hall resistivity ($\varrho_{xy}$) exhibits
other features (Fig.~\ref{rxy}), which are also in good
quantitative agreement with experiment.
Secondly, we find that the negative Hall resistivity is absent for
infinitely steep antidots (Sinai billiard) where only a few
channeling trajectories exist, whereas the rebound mechanism would be
compatible with steep walls.
Furthermore the quenching mechanism is not an edge but a bulk effect as it
shows up even in the Hall {\em resistivity} and not only in the {\em
resistance}.

In microjunctions and lateral
surface superlattices the dynamics of electron wave packets approaches the
classical limit as the Fermi wavelength is smaller than the typical
length scale of the lateral structure.
We model the antidot superlattice with lattice constant $a$ by the
continuous potential \cite{prl} (Fig.~\ref{traj} right inset)
\begin{equation}
U(x,y)=U_{0}[\cos(\pi x/a)\cos(\pi y/a)]^{\beta} ,
\label{eq:pot}
\end{equation}
with $\beta = 4$.
The prefactor $U_{0}$ of the potential is chosen such that the ratio
of the dot diameter at the Fermi energy $\varepsilon_{F}$ to the distance
of adjacent dots is one-third, as in the experiments \cite{weiss}.
The classical approximation for the dynamics of an electron wave
packet in the potential $U(x,y)$ and a perpendicular magnetic field
${\bf B} \parallel {\bf e}_{z}$ is then given by 4 coupled differential
equations, which in
scaled variables ($x \rightarrow x/a$, $ y \rightarrow y/a$,
$ U \rightarrow U/\varepsilon_{F}$, $ t \rightarrow t/t_{0}$, with
$t_{0}=(\varepsilon_{F}/ma^{2})^{1/2}$) read
\begin{eqnarray}
\dot{x} = v_{x} & \;\; & \dot{v}_{x}  =  \hphantom{-}2\sqrt{2}(B/B_{0})v_{y} -
\partial U/\partial x  \label{eq:motion} \\
\dot{y} = v_{y} & \;\; & \dot{v}_{y}  =  -2\sqrt{2}(B/B_{0})v_{x} - \partial
U/\partial y \nonumber .
\end{eqnarray}
A chaotic trajectory for small magnetic field
$B/B_{0}=0.1$ is shown in Fig.~\ref{traj}, where
$B_{0}=2(2m\varepsilon_{F})^{1/2}/(ea)$ corresponds with a free
cyclotron radius of $a/2$ of an electron with effective mass $m$. The
trajectory follows one of the channels in the potential interrupted
by local chaotic motion. Thus the free cyclotron motion in a magnetic
field is destroyed by the periodic potential and the electron is forced
into long episodes of linear motion.
This is visualized in more detail in the left inset of Fig.~\ref{traj}.

We investigate the motion in phase space $(x,y,v_{x},v_{y})$
for $B \neq 0$ and $E=0$, by means of Poincar\'e surfaces
of section $(x,v_{x})$ at $y({\rm mod} 1)=0$ for various initial conditions.
In Fig.~\ref{ps} for $v_{y}>0$ (left) and $v_{y}<0$ (right), this
section displays a sea of chaotic motion and an island of regular
motion which represents a tube of trajectories channeling forever in the
$y$-direction.
It contains periodic and quasiperiodic trajectories
lying on cylindrical KAM-surfaces \cite{KAM} around a central trajectory
in phase space. They arise from nonlinear resonances between the degrees of
freedom and are represented by closed loops in the
Poincar\'e surface of section.
In the neighbourhood of such a tube of regular channeling trajectories,
a chaotic trajectory can be trapped in an infinite hierarchy of
permeable barriers, so-called cantori (not visible in Fig.~\ref{ps}).
This appears to be the origin of the long channeling episodes of the
chaotic trajectory in Fig.~\ref{traj}. They are associated with a long-time
tail as in similar systems which we previously studied
\cite{geisel87,geisel90}. Here they will turn out to be essential for
the negative Hall effect. We note that the tubes of regular motion in
{\em both} directions ($v_{y}>0$ and $v_{y}<0$) as well as
surrounding chaotic channeling trajectories still exist in
a weak electric field {\bf E} $\parallel{\bf e}_{x}$. This means that
many trajectories move in opposite direction to the free
${\bf E \times B}$-drift and qualitatively explains the quenching of the
Hall effect.
For a more detailed account we now apply a linear response theory
including regular as well as chaotic trajectories.

We therefore divide phase space into regions of chaotic and
regular motion corresponding to Fig.~\ref{ps} and write the
conductivity $\mbox{\boldmath $\sigma$}$ as
\begin{equation}
\mbox{\boldmath $\sigma$} = p_{r} \mbox{\boldmath $\sigma$}^r + p_{c}
\mbox{\boldmath $\sigma$}^{c} ,
\end{equation}
where $p_r$ and $p_c$ are the portions of phase space volume occupied by
regular and chaotic trajectories, respectively.  Drude-type considerations
for the straight regular trajectories in the channels lead to longitudinal
conductivities $\sigma_{xx}^{r}=\sigma_{yy}^{r}=p_{r}n \tau e^{2} / 2m$,
where $n$ is the electron density
and $\tau$ is the mean elastic scattering time, and to vanishing transversal
conductivities $\sigma_{xy}^{r} = \sigma_{yx}^{r}=0$.
Combining this with the conductivity of chaotic trajectories as
derived in Ref.~\cite{prl} using linear response theory yields
\begin{equation}
\sigma_{ij} = p_{r} \sigma_{ij}^{r} \; +\; (1-p_{r}) \sigma_{0}
\int_{0}^{\infty}\;e^{-\case{t}/{\tau }} \;C_{ij}(t)\;dt ,
\label{eq:transp}
\end{equation}
where $C_{ij}(t) = \langle v_{i} (t) v_{j} (0) \rangle $ is the correlation
function of the {\em unperturbed} chaotic trajectories
(i.e.\ neither perturbed by an electric
field nor by impurity scattering ) and the factor $exp(-t/\tau )$ describes
impurity scattering \cite{remark1}.
The numerical analysis involves the determination of $p_r$ as the volume
of the outermost invariant KAM-surface enclosing the regions of
regular motion divided
by the total phase space volume. As these regular regions are small, however,
the main contribution to the conductivities calculated according to
Eq.~(\ref{eq:transp}) stems from chaotic trajectories, whose
correlation functions $C_{ij}(t)$ were determined from numerical
simulations. The resulting Hall resistivity $\varrho_{xy}$
(Fig.~\ref{rxy}) shows a quenching of the Hall effect and even a
negative Hall resistivity as in the experiment \cite{weiss}.
The negative Hall resistance in antidot superlattices thus is a bulk
effect due to chaotic electron trajectories.

We now analyze the dynamical features leading to a negative Hall effect
at small magnetic fields.
The sign of the Hall resistivity $\varrho_{xy} = \sigma_{yx}
/(\sigma_{xx}^2+\sigma_{yx}^2)$ ( with
$\sigma_{xy} = -\sigma_{yx} $) is given by the sign of $\sigma_{yx} $,
which according to Eq.~(\ref{eq:transp})
is determined essentially by the correlation function $C_{yx}(t) $
of the chaotic trajectories.
Figure~\ref{corr} shows the time dependence of $C_{yx}(t) $ for $B/B_{0}=0.05$
where a negative Hall resistivity was found. Its initial increase still
follows the free electron case but for later times there is a strong
negative peak and a {\em negative algebraic long time tail}.
This tail is responsible for the negative Hall effect as it yields a
dominance of negative contributions in the integral over time in
Eq.~(\ref{eq:transp}). Note that the mean elastic scattering time
is $\tau \approx 25$ in the experiment \cite{weiss}.
For larger magnetic fields ($B/B_{0}\geq 0.12$, see dotted line in
Fig.~\ref{corr}) the tail disappeares and $C_{yx}(t)$ exhibits an
oscillatory decay where positive contributions to $\sigma_{yx}$
progressively outweigh the negative ones. The oscillatory decay
for $B/B_{0}\geq 0.12$ is caused by chaotic trajectories that roughly
follow {\em cyclotron-type orbits} around 1 or 4 antidots for a limited time
and were shown to be responsible for pronounced peaks in the
magnetoresistance $\varrho_{xx}(B)$ (Ref.~\cite{prl} and dotted line in the
inset of Fig.~\ref{rxy}). Many rebound trajectories \cite{ford,been,baranger}
would have a similar curvature and would be expected in this magnetic
field range ($B/B_{0}\approx 0.5$), yet they do not lead to a quenched
or negative Hall effect. Instead one finds a prominent Hall plateau, which
corresponds well with the experimental results
\cite{weiss,weisspriv,ensslinmaut}. In the experiment of Ref.~\cite{weiss},
e.g.\ $B_{0}\approx 0.55$T.

The long-time tail stems from chaotic {\em channeling trajectories},
which exhibit long episodes of straight motion (Fig.~\ref{traj})
and dominate the dynamics below $B/B_{0}=0.12$.
The distribution $\psi(\ell)$ of channeling lengths decays algebraically
$\psi(\ell) \propto \ell^{-\nu}$ where $\nu$ fluctuates
around $\nu=3$. Long-time tails seem to be typical for chaotic Hamiltonian
systems and are believed to be related to cantorus hierarchies (see e.g.\
Ref.~\cite{meiss,geisel87}) as described in the context of Fig.~\ref{ps},
but their origins are still not understood definitively. In the present
case, after moving in a channel e.g.\ along the $x$-axis
for a while (at least for 3 unit cells)
the trajectories tend to leave the channel in the direction {\em opposite}
to the Lorentz force $-e{\bf v}\times{\bf B}$ (which points to their left).
For example at $B/B_{0}=0.05$  a majority of about
$65\% $ of the trajectories show this anomaly (see Fig.~\ref{corr} inset)
and thereby yield the negative sign of the long-time tail.
Intuitively the experimental situation can be
imagined by an applied current transported mainly by channeled (chaotic)
electrons moving in positive $x$-direction.
Since they tend to leave the channel
in negative $y$-direction they build up a negative Hall voltage.

The role of the chaotic channeling trajectories for the long-time tail and
the negative Hall effect can be described in more detail.
For times $t\gg a/v_{F}$ and small magnetic fields the decaying correlation
function $C_{yx}(t)$ is determined by long channeling episodes and
can be expressed analytically in terms of the length distribution $\psi(\ell)$
by \cite{tbp}
\begin{equation}
C_{yx}(t) \sim \frac{2\alpha v_{F}^2}{\langle \ell\rangle}
\int_{v_{F}t}^{\infty}\; \psi(\ell)\; \left[ p_{+}(\ell)-p_{-}(\ell)
\right] \;d\ell,
\label{psi2c}
\end{equation}
where $v_{F}$ is the Fermi velocity, $\alpha v_{F}^2$ is the mean absolute
value of the contribution to the correlation function of individual channeling
episodes, and $\langle \ell\rangle$ is
the mean length of these episodes. $p_{+}(\ell)$ and $p_{-}(\ell)$
are the conditional probabilities that a trajectory trapped in a channeling
episode of length $\ell$ leaves the channel in {\em positive} or
{\em negative} direction with respect to the Lorentz force.
The asymptotic decay of $C_{yx}(t)$ obtained form Eq.~(\ref{psi2c}) is
indicated by the dashed line in Fig.~\ref{corr}. Here $\alpha$ and
$p_{+}(\ell)-p_{-}(\ell)$ were obtained approximately from geometrical
considerations, while for the length distribution $\psi(\ell)$ we used the
numerical data. The long-time tail of $C_{yx}(t)$ is relevant for the
negative Hall resistivity up to times $t\approx \tau$,
i.e.\ $\ell \approx 30$.
The dashed line in Fig.~\ref{corr} demonstrates that it is caused by the
distribution $\psi(\ell)$ of chaotic channeling lengths.

We emphasize that the results presented here do not depend sensitively on a
particular choice for the periodic potential. While for an infinitely steep
antidot potential the effects are absent, realistic potentials with a small
degree of softness produce a negative tail in $C_{yx}(t)$ (we have
checked this e.g.\ for $\beta= 20$ in Eq.~(\ref{eq:pot}).
For the sample used in Ref.~\cite{weiss} $\beta=4$ seems realistic \cite{prl}).
The strength of the tail determines whether Eq.~(\ref{eq:transp}) yields
a negative Hall resistivity or only  quenching. Furthermore irregularities
of the potential lead to a reduction of the tail. (For a distorted potential
with $\pm 5 \%$ disorder in the antidot positions we still found a negative
Hall resistivity.) This might explain why in some experimental situations
quenching and in others a negative Hall effect was
reported \cite{weiss,ensslin}.

We acknowledge motivating discussions with R.~R.~Gerhardts, K.~v.~Klitzing,
and D.~Weiss. This work was supported by the Deutsche Forschungsgemeinschaft.
RK thanks Walter Kohn for the hospitality in Santa Barbara
and for partial support by the NSF under DMR90-01502.

\begin{figure}
\caption{
A chaotic trajectory in the antidot potential (right inset) most of the time
shows straight paths along a channel even though a
perpendicular magnetic field ($B/B_{0}=0.1$)
is applied. A magnification of such a
channeled trajectory in positive and negative $y$-direction (left inset)
shows an interplay of left turns due to the magnetic field
and of right turns due to deflections from the antidot potential.
\label{traj}}
\end{figure}

\begin{figure}
\caption{
Poincar\'e surfaces of section
at $y({\rm mod} 1)=0$ for $v_y>0$ (left) and $v_y<0$ (right) for a weak
magnetic field $B/B_{0}=0.1$. The closed curves in the center represent
tubes of regular motion intersecting the plane perpendicularly in both
directions. The quenching of the Hall effect is related to the survival
of these tubes in {\em both} directions ($v_{y}>0$ and $v_{y}<0$)
when a weak electric field {\bf E} $\parallel{\bf e}_{x}$ is
applied.
\label{ps}}
\end{figure}

\begin{figure}
\caption{
Hall resistivity $\varrho_{xy}$ versus magnetic field $B/B_0$
showing a negative Hall effect for $B/B_0<0.12$ calculated
from Eq.~(\protect\ref{eq:transp}). Above $B/B_0=0.12$ the inset
shows a prominent plateau in $\varrho_{xy}$ for $0.4 < B/B_{0} <0.9$
(solid line) caused by chaotic trajectories revolving around 1 or 4 antidots
for a while (the dotted line shows $\varrho_{xx}$ after
Ref.~\protect\cite{prl} for comparison).
The small plateau in $\varrho_{xy}$ and the shoulder in $\varrho_{xx}$
at $B/B_0 \approx 0.25$ are due to chaotic trajectories enclosing
2 antidots. (The $\varrho_{xx}$ and $\varrho_{xy}$ scales are not
identical.)
\label{rxy}}
\end{figure}

\begin{figure}
\caption{
The velocity correlation function $C_{yx}(t)$ for $B/B_0=0.05$ (solid line)
exhibits a large negative peak and a negative algebraic tail. Both are caused
by channeled chaotic trajectories (inset), which tend to leave the channel
in a direction opposite to the Lorentz force $-e{\bf v}\times{\bf B}$ and thus
give a negative contribution to the correlation.
According to Eq.~(\protect\ref{eq:transp}) the
area enclosed under $C_{yx}(t)$ up
to the elastic scattering time $\tau=25$ \protect\cite{weiss}
is a measure for the
Hall conductivity. --- Above $B/B_0=0.12$ other chaotic
trajectories dominate, which revolve
around $1$ or $4$ antidots and thereby produce oscillations in $C_{yx}(t)$
(dotted line for $B/B_{0}=0.15$).
The dashed line shows how the length distribution $\psi(\ell)$ according
to Eq.~(\protect\ref{psi2c}) gives rise to the tail for $B/B_0=0.05$.
\label{corr}}
\end{figure}


\begin{references}
\bibitem{remark1}
The exponential factor describes elastic impurity scattering only in first
order. This is sufficient however as the effect of impurity scattering
on the correlation decay is small compared to the effect of chaotic scattering.
\end{references}
\end{document}